\documentclass[%
 aip,
 amsmath,amssymb,
 reprint,%
]{revtex4-2}

\usepackage{graphicx}
\usepackage{dcolumn}
\usepackage{bm}

\usepackage[utf8]{inputenc}
\usepackage[T1]{fontenc}
\usepackage{mathptmx}
\usepackage{etoolbox}

\makeatletter
\def\@email#1#2{%
 \endgroup
 \patchcmd{\titleblock@produce}
  {\frontmatter@RRAPformat}
  {\frontmatter@RRAPformat{\produce@RRAP{*#1\href{mailto:#2}{#2}}}\frontmatter@RRAPformat}
  {}{}
}%
\makeatother
\begin{document}

\preprint{AIP/123-QED}

\title[Local work function of GNRs/Au]{Local work function on Graphene Nanoribbons and on the Au(111) herringbone reconstruction}
\author{D. Rothhardt}
\affiliation{Institute of Physics and Astronomy, University of Potsdam, Karl-Liebknecht-Str. 24-25, 14476 Potsdam-Golm, Germany} 

\author{A. Kimouche}
\email{amina.kimouche@uni-potsdam.de}
\affiliation{Institute of Physics and Astronomy, University of Potsdam, Karl-Liebknecht-Str. 24-25, 14476 Potsdam-Golm, Germany}

\author{T. Klamroth}
\affiliation{Institute of Chemistry, University of Potsdam, Karl-Liebknecht-Str. 24-25, 14476 Potsdam-Golm, Germany}

\author{R. Hoffmann-Vogel}
\email{Regina.Hoffmann-Vogel@uni-potsdam.de}
\affiliation{Institute of Physics and Astronomy, University of Potsdam, Karl-Liebknecht-Str. 24-25, 14476 Potsdam-Golm, Germany}

\date{\today}

\begin{abstract}
Graphene nanoribbons show exciting electronic properties related to the exotic nature of the charge carriers and to local confinement as well as atomic-scale structural details.
The local work function provides evidence for such structural, electronic and chemical variations at surfaces. Kelvin prove force microscopy (KPFM) can be used to measure the local contact potential difference (LCPD) between a probe tip and a surface, related to the work function. Here we use this technique to map the LCPD of graphene nanoribbons grown on a Au(111) substrate. The LCPD data shows charge transfer between the graphene nanoribbons and the gold substrate. Our results are corroborated with density functional theory calculations which verify that the maps reflect from the doping of nanoribbons. Our results help to understand the relation between atomic structures and electronic properties.
\end{abstract}

\maketitle

Graphene's electronic properties are determined by its two-dimensionality as well as by its semimetallic gapless conical band structure \cite{castroneto09p1}. Its electronic behavior depends strongly on the location of the Fermi level with respect to the Dirac point, the center of the Dirac cones \cite{gierz08p1}. This position can be tuned by gating\cite{koch12p1} or by doping, e.g. n-doping on SiC\cite{kedzierski08p1,gu07p1} and p-doping by Bi, Sb and Au\cite{gierz08p1}. Confining graphene to nanostructures \cite{li08p1,narita15p1} and in particular to graphene nanoribbons (GNRs), few-nm-wide stripes of graphene, opens additional possibilities of tuning the electronic properties by creating quantum-confined states\cite{hamalainen11p1} and opening a size-dependent energy gap\cite{li08p1,ritter09p1}. Also for GNRs the electronic properties are strongly influenced by charge transfer between the substrate and the GNR \cite{vanin10p1}. Additionally, the chemical state of GNR's edges has a strong influence \cite{wagner13p1}. GNRs can be produced in an atomically precise ultra-high vacuum compatible way using on-surface synthesis \cite{cai10p1}. This synthesis is well-known to be performed on the coinage metals, Cu, Ag and Au containing a large density of electrons. 

As a basis to study these unique electronic properties a suitable method to study the charge transfer, i.e. the work function, between a GNR and a metal substrate at the atomic scale is needed. The work function can provide evidence for structural, electronic and chemical variations at surfaces \cite{melitz11p1}. Kelvin probe force microscopy (KPFM), a method derived from scanning force microscopy (SFM), allows to study the local work function of a sample with great accuracy and with atomic resolution \cite{nonnenmacher91p1,kikukawa96p1,jacobs98p1,barth11p1,sadewasser09p1,perez16p1}.
In KPFM a voltage is applied to the tip in order to compensate electrostatic forces occurring between tip and sample. Such electrostatic forces arise from the different position of the Fermi level in tip and sample that gives rise to charge transfer. In KPFM the forces are measured by SFM during image acquisition \cite{giessibl03p1,garcia02p1}. In this way an image of the local contact potential difference between tip and sample is obtained as has been also shown for molecules and molecular layers \cite{mohn12p1,barth11p1,zerweck05p1,neff14p1}. 

Here, we study the charge transfer and the local work function for graphene nanoribbons fabricated by on-surface synthesis on Au(111). We obtain local variations in contact potential differences on the Au(111) herringbone reconstruction that we relate to the different crystallographic structure of the Au in the different regions of the reconstruction. The GNRs can be clearly discerned from the reconstruction through their topography but also through their contact potential difference. GNR has a contact potential that is about 100 meV smaller than that of a Au indicating that GNRs are charged positively compared to Au. This is confirmed by calculations.

\begin{figure}
   \includegraphics[width=0.8\linewidth,angle=0,clip]{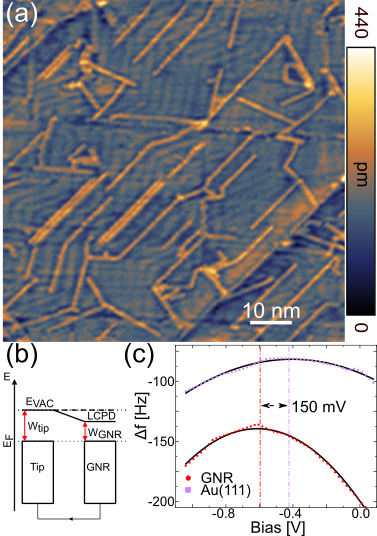}
     \caption{a) Topography image of GNRs on Au, $f_{0}=158$~kHz, $c_{L}=41$ N/m, $A=3$ nm, $Q=31000$ and $\Delta f=-28$ Hz. b) In KPFM local variations in contact potential (CPD) can be measured, by applying a voltage between the sample and the AFM tip so that the electric field caused by the CPD is compensated. c) $\Delta f(V)$ spectra along with their second-order polynomial fit measured on GNR and Au. The dashed-dotted lines indicates the shift of the work function.}
     \label{fig_1}
\end{figure}

\begin{figure}
   \includegraphics[width=0.8\linewidth,angle=0,clip]{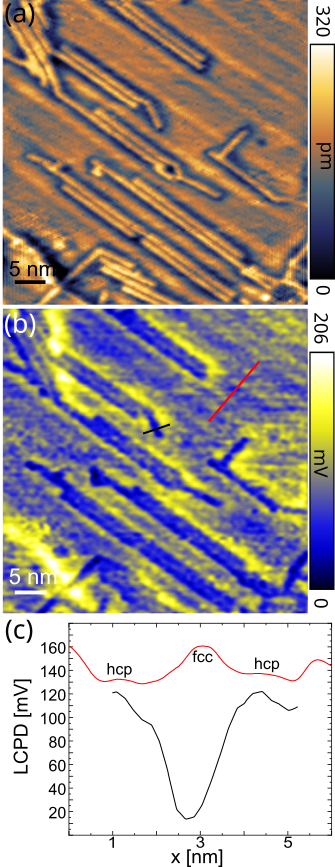}
     \caption{a) Topography of GNR's and the Au(111) herringbone reconstruction, $f_0=158$ kHz, $c_L=41$ N/m, $A=5$~ nm, $Q=31000$ and $\Delta f=-28$ Hz. b) LCPD image recorded simultaneously with the topographic image. c) LCPD line profiles taken across a ribbon (black) and herringbones (red) shown in b).}
     \label{fig_2}
\end{figure}

A topographic image of GNRs on Au surface is shown in Fig.~\ref{fig_1} a). While most GNRs are attached to gold step edges or to other ribbons, we additionally observe isolated individual ribbons. GNRs are aligned along three well-defined orientations with respect to the high symmetry Au(111) directions.
When the tip and the GNR are brought close together such that charges can equilibrate, their Fermi level align, accompanied by an electron flow to the Au and the GNR is charged leading to a local contact potential difference (Fig.~\ref{fig_1} b). 
In Fig.~\ref{fig_1} c), $\Delta f(V)$ curves measured above a GNR and Au are shown. The maxima of the parabola fitted to the measured data yield the difference of the LCPD values: $\Delta V$ = 150 mV. As a result, this shift is explained by p-doping from the substrate due to intrinsic charge transfer across the interface, also seen in bulk graphene on a gold substrate \cite{giovannetti08p1,wofford12p1,leicht14p1,gierz08p1}.
Fig.~\ref{fig_2} shows a topographic image of GNRs on Au(111) surface and its associated Kelvin voltage map. The GNR and the Au(111) herringbone structures are observed. In the LCPD image (Fig.~\ref{fig_2} b), GNRs appear as dark stripes on the Au(111) herringbone reconstruction. The compensation of the electrostatic potential allow us to determine ribbon heights with high accuracy. A line profile taken across a ribbon is shown in Fig.~\ref{fig_2} c) together with a line profile taken across a herringbone reconstruction. Clearly the two line profiles are different, showing that our measurements are clearly capable of unambiguously determining the LCPD of a GNR.

First, we analyze the work function differences we find on the Au herringbone reconstruction. The measured contact potential difference across the herringbone reconstruction varies between 10 and 30 mV (Fig.~\ref{fig_2} c). This shows how small differences in the coordination and the local electronic structure give rise to changes in the local potential of different surface regions. This potential is related to a change in local surface reactivity. Theory predicts that indeed the local coordination environment results in a small variation of the reactivity across different surface sites using hydrogen as a probe atom with an interaction energy difference up to 45 meV, see Ref. \cite{hanke13p1}.\\ 
Our findings are also in agreement with STM studies, where the reconstruction of Au leads to a modulated electron density with estimated electronic energy potential to be 25 meV between fcc and hcp regions \cite{chen98p1}. We conclude that these energy differences are largely related to the charge accumulated at these areas and an electrostatic field related to this charge resulting in an enhanced chemical reactivity at the fcc sites. Our KPFM map presents a robust method to probe electron potentials on atomic length scales quantitatively.\\
Near the edges of the GNR, on the Au(111) we can see an area of modified work function of Au (Fig. \ref{fig_2} b). We relate this area to the screening in metallic Au(111) of electrostatic fields originating from the GNR. These electrostatic fields have been calculated previously and are known to be strongest near the edges and corners of a GNR \cite{wang10p1}.

We calculate the local work function $\Phi(\underline{r})$ from the Hartree potential $V_{\text{eff}}(\underline{r})$ corrected by the Fermi energy, i.e. $\Phi(\underline{r}) = V_{\text{eff}}(\underline{r}) - E_\text{Fermi}$ as done in Ref.~\cite{perez16p1}. We use different constant values of $z$ for the LCPD maps.  $\underline z$ is parallel to $\underline c$, $\underline x$ parallel to $\underline a$, i.e. the long axis of the GNR and $\underline y$ to $\underline b$. The LCPD maps are derived form
\begin{equation}
  \Phi^{(s)}(x,y)= \Phi(x,y,z_\text{surf} +s) \qquad ,
\end{equation}
where $z_\text{surf}$ is given by the sum of the $z$ coordinate of the uppermost carbon atom of the GNR and the van der Waals radius of carbon, i.e., 1.7 \AA{}. $s$ is varied from 0 \AA{} to 4.5 \AA{}. Details about the density functional theory (DFT) calculations performed in this work are given in the supporting information.

\begin{figure}
   \includegraphics[width=0.8\linewidth,angle=0,clip]{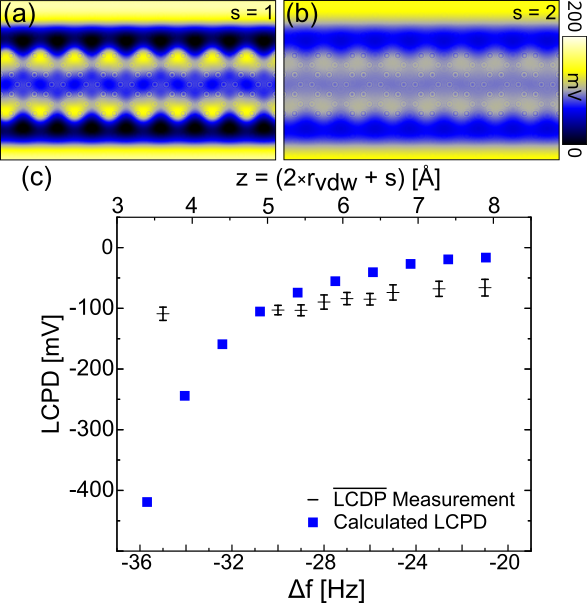}
     \caption{a,b) Calculated LCPD maps at distance of  $s$ = 1 and 2 \AA{} from the GNR plane. c) Mean LCPD values on GNR/Au versus frequency shift in comparison with calculated LCPD.}
     \label{fig_3}
\end{figure}

Fig.~\ref{fig_3} a and b are LCPD maps calculated from the Hartree potential of GNR/Au(111). In the calculations we did not represent the Au herringbone reconstruction, because this is computationally very demanding. With decreasing tip-sample distance, the emergence of submolecular structure is observed within the GNR (see Supporting Information) and the GNR appears as a featureless depression. The intramolecular contrast is found to increase with decreasing $z$. 

To obtain a more detailed understanding of the charge transfer and for comparison with calculations, we performed measurements on several different positions on the surface, as depicted in Fig.~\ref{fig_3} c). With this statistical approach (over 257 LCPD line scans) the influence of the tip and sample microstructures on the resulting overall values is minimized. Additionally, the $\overline{LCPD}$ are measured with respect to the reference LCPD recorded on the Au(111) surface. In Fig.~\ref{fig_3} c) the $\overline{LCPD}$ value exhibits slow increase with frequencies. Depending on the selected frequency shift, the $\overline{LCPD}$ varies from 60 mV for -21 Hz to 120 mV for -35 Hz. In other words, a higher work function is measured above the Au surface as compared to the GNR, which indicates that the electrons prefer to reside in the Au metal. As a result, the predicted hole doping is confirmed.

For comparison, we also plot the calculated LCPD values in the same graph, Fig.~\ref{fig_3} c. At distance of closest approach greater than $4$ \AA{}, the calculated LCPD values are comparable to the measured ones. Instead, at close tip-sample distances, the calculated LCPD values decrease significantly, leading to an enhancement of the sensitivity to the short-range field generated by charges. 

As the calculations agree well with the measurement, it can be deduced that the measured LCPD values were obtained with a tip-sample distance ranging from s = 2.0 \AA{} to 4 \AA{}. The above observations reveal a clear dependence of LCPD signal with tip-sample distance, attributed to variation in electrostatic forces \cite{schneider20p1}, in agreement with previous studies \cite{kawai10p1,okamoto03p1,krok08p1,enevoldsen08p1}. This finding is consistent with the interpretation of the distance dependence of the LCPD signal, thus confirming that our measured LCPD signal reflects the properties of tip apex surface interactions.

In summary, we have imaged graphene nanoribbons and the Au(111) herringbone reconstruction using KPFM. We have shown that KPFM yields a charge distribution on the GNRs and herringbones. LCPD studies confirm the p-type doping of the GNRs on the Au substrate. We found that the measured LCPD values exhibits a slow decrease with frequency, whereas the calculated LCPD exhibits a strong dependency on the tip-sample distance. Our results highlight the potential of Kelvin probe force microscopy to simultaneously study structural and electronic properties of GNRs and to map the electron potential in Au. Our findings show the capability of KPFM as a useful tool for observing the electronic properties in nanoelectronics.\\

See the supplementary material for additional information about the experimental procedures and the DFT calculations.

\begin{acknowledgments}
We thank T. Beitz, P. P. Schmidt and B. Weinschenk for help with the experiments.
\end{acknowledgments}

\section*{AUTHOR DECLARATIONS}
\section*{Declaration of interest}
The authors have no conflicts of interest to disclose.


\section*{Data Availability Statement}

The data that support the findings of this study are available within the article and its supplementary material.


\nocite{*}

\end{document}